\documentclass[runningheads,a4paper, 11pt]{llncs}

\usepackage{graphicx}
\usepackage{epstopdf}
\usepackage{setspace}
\usepackage{amsfonts}
\usepackage{amssymb}
\usepackage{stmaryrd}
\usepackage{bbding}
\usepackage{amssymb}
\usepackage{algorithm}
\usepackage{paralist}
\usepackage{mdwlist}
\usepackage{algpseudocode}
\usepackage{multicol}
\usepackage{lipsum}
\usepackage{subfigure}
\usepackage{booktabs}
\usepackage{verbatim}
\usepackage{graphicx}
\usepackage{epstopdf}
\usepackage{amsmath}
\newtheorem{mydef}{Definition}
\usepackage{multicol}
\usepackage{appendix}

\begin{document}

\title{PP-MCSA: Privacy Preserving Multi-Channel Double Spectrum Auction}

%
%
%

\author{Zhili Chen\inst{1} \and
Sheng Chen\inst{1} \and
Hong Zhong\inst{1} \and
Lin Chen\inst{2} \and
Miaomiao Tian\inst{1}}
\authorrunning{Z. Chen et al.}
%
\institute{School of Computer Science and Technology, Anhui University, 230601 Hefei, China \and
Lab. Recherche Informatique (LRI-CNRS UMR 8623), Univ. Paris-Sud, 91405 Orsay, France
\email{Email: zlchen@ahu.edu.cn, shengchen9403@gmail.com, zhongh@mail.ustc.edu.cn, chen@lri.fr, mtian@ahu.edu.cn}
}

\maketitle              
\begin{abstract}
Auction is widely regarded as an effective way in dynamic spectrum redistribution. Recently, considerable research efforts have been devoted to designing privacy-preserving spectrum auctions in a variety of auction settings. However, none of existing work has addressed the privacy issue in the most generic scenario, double spectrum auctions where each seller sells multiple channels and each buyer buys multiple channels. To fill this gap, in this paper we propose PP-MCSA, a \underline{P}rivacy  \underline{P}reserving mechanism for \underline{M}ulti-\underline{C}hannel double \underline{S}pectrum \underline{A}uctions. Technically, by leveraging garbled circuits, we manage to protect the privacy of both sellers' requests and buyers' bids in multi-channel double spectrum auctions. As far as we know, PP-MCSA is the first privacy-preserving solution for multi-channel double spectrum auctions. We further theoretically demonstrate the privacy guarantee of PP-MCSA, and extensively evaluate its performance via experiments. Experimental results show that PP-MCSA incurs only moderate communication and computation overhead.

\end{abstract}

\section{Introduction}

Today, more and more emerging wireless technologies, such as Wifi, 4G, are penetrating into our daily work and life. At the same time, the traditional static and rigid spectrum allocation scheme renders the utilization of radio spectrum severely inefficient and unbalanced. According to the survey {\cite{valenta2010survey}}, many statically allocated spectrum channels are left idle by their current owners, exaggerating the gap between the ever-increasing spectrum demand of wireless services and the spectrum scarcity. Therefore, to improve and balance spectrum utilization, dynamic spectrum redistribution has been advocated to reallocate spectrum among primary and secondary users.

Spectrum auction is widely regarded as an effective way in dynamic spectrum redistribution. A large body of existing studies are focused on designing truthful spectrum auctions, where the auctioneer is assumed to be trusted, and bidders are stimulated to reveal their true valuations of spectrum channels. However, in many practical scenarios, the auctioneer is by nature self-interested and not trusted. It may disclose the true valuations of bidders, which may cause serious privacy vulnerabilities \cite{11panm}. For example, a dishonest auctioneer may take advantage of learning the bidders' bids, and then tamper with the auction results so as to increase its own profit. Or the auctioneer may sell bidders' historical bids for profit. Therefore, privacy preservation is critical in spectrum auctions.

There has been significant research attention on privacy preserving auctions, such as~\cite{99naorm,04yokoom,02pengk}. These schemes do not consider spectrum reusability, and thus cannot be applied in spectrum auctions. Recently, a handful of propositions addressed privacy issues in spectrum auctions, such as \cite{14huangh,13huangq,11panm}, but most of them focus on protecting privacy for single-sided spectrum auctions. Only a few solutions such as \cite{14chenz} and \cite{chen2017secure}, provide secure designs for double spectrum auctions. However, they assume that in the auction each seller sells only one spectrum channel and each buyer buys only one spectrum channel. Such \emph{one-channel} assumption makes the problem much more tractable, but leaves open the most generic and practical version, the double spectrum auctions, involving
multiple spectrum sellers selling multiple channels to multiple
buyers~\cite{13chenz}.

\begin{figure}
  \centering
  \includegraphics[width=0.6\linewidth]{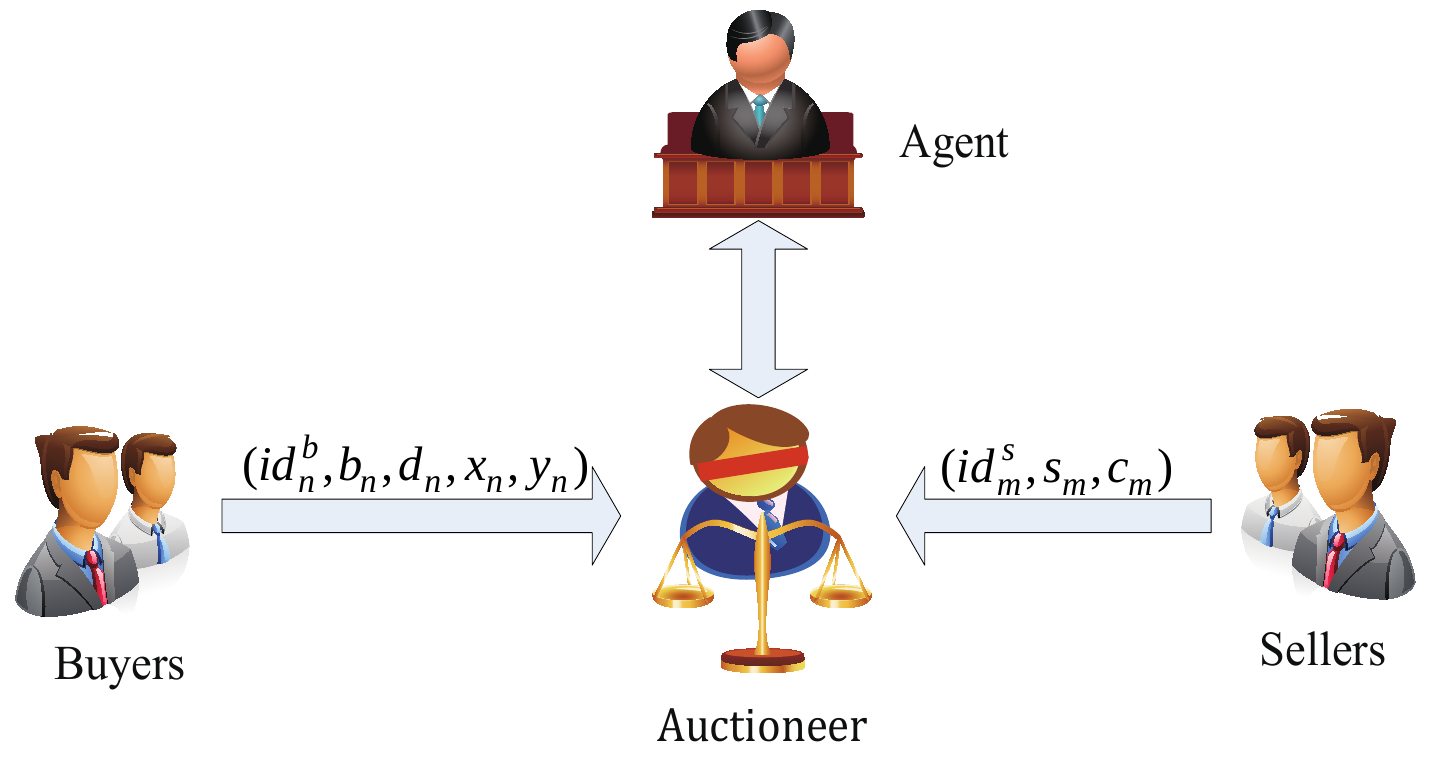}
  \caption{Privacy-preserving auction framework for PP-MCSA}\label{fig:AuctionFramework}
\end{figure}

To fill this gap, in this paper, we propose PP-MCSA, a \underline{P}rivacy-\underline{P}reserving mechanism for \underline{M}ulti-\underline{C}hannel double \underline{S}pectrum \underline{A}uctions. Specifically, we manage to protect both sellers' request privacy and buyers' bid privacy for the double spectrum auction mechanism True-MCSA \cite{13chenz} that supports multi-channel auctions. To preserve privacy, we introduce in the auction framework of PP-MCSA a third party, namely an agent, who cooperates with the auctioneer to perform secure auction computations, as shown in Fig.~\ref{fig:AuctionFramework}. In such a framework, each seller $m$ submits its request value $s_m$ (i.e., the lowest per-channel selling price) and request number $c_m$ (i.e., the number of selling channels) to the auctioneer. Similarly, each buyer $n$ does the same thing with its bid value $b_n$ (i.e., the highest per-channel buying price) and bid number (i.e., the number of buying channels). All submissions are appropriately encrypted such that all sensitive information (i.e. request values, bid values and bid numbers) are protected from either the auctioneer or the agent, but can be securely retrieved and computed with the cooperation between the two parties. Therefore, as long as the auctioneer and the agent do not collude with each other (Note that this assumption is essentially necessary, otherwise the privacy cannot be achieved.), PP-MCSA leaks nothing about the sensitive information to anyone except what can be revealed from the published auction outcome.

We list our main contributions as follows:
\begin{compactitem}
  \item We propose the first privacy-preserving and practical multi-channel double spectrum auction mechanism by combing public-key encryptions and garbled circuits, filling the research gap that there is no privacy consideration in multi-channel double spectrum auctions before.
  \item We design and optimize data-oblivious algorithms for multi-channel double spectrum auction mechanism True-MCSA, which is rather complex in auction logic, and address both the privacy and efficiency challenges. 
  \item We fully implement PP-MCSA, and conduct extensive experiments to evaluate its computation and communication overhead.
\end{compactitem}


The reminder of this paper is structured as follows. Section~\ref{sec:relatedwork} briefly reviews related work. In Section~\ref{sec:Problem Statement}, the underlying mechanism is introduced and the privacy goal is given. We describe the design challenges and rationale in Section~\ref{sec:challenges}, and present the detailed design of PP-MCSA and prove its privacy in Section \ref{sec:DesignDetail}. In Section
\ref{sec:experiment}, we implement PP-MCSA, and evaluate its performance in terms of computation and
communication overheads. Finally, the paper is concluded in Section
\ref{sec:conclusion}.

\section{Related Work}\label{sec:relatedwork}
In this section, we briefly review the existing works on privacy-preserving auction design, and distinguish our work from the existing ones.
\subsection{Spectrum Auction}
Spectrum auctions are widely used to redistribute spectrum. In the past few years, many researches have focused on designing truthful spectrum auctions. For example, Zhou et al. put forward TRUST{\cite{09zhoux}}, the first truthful double spectrum auction framework exploiting spectrum reusability. Chen et al. proposed the first truthful single-sided auction mechanism TAMES{\cite{14cheny}} for heterogeneous spectrum auctions, which allows buyers to freely bid their different preferences to heterogenous spectrum channels. Later, Feng et al. presented the first double auction mechanism for heterogeneous spectrum transaction {\cite{feng2012tahes}}. Chen et al. proposed the first double multi-channel spectrum auction scheme, True-MCSA{\cite{13chenz}}. However, all the above studies did not address the privacy preservation issues.
\subsection{Privacy-preserving Spectrum Auction}
In the past decade, there have been a great number of schemes for privacy-preserving auctions {\cite{99naorm}}{\cite{04yokoom}}{\cite{02pengk}}. These schemes were originally designed for traditional goods (e.g., painting, stamps), where each commodity can only be allocated to one bidder. Unfortunately, when directly applied to spectrum
auctions, they suffer severe under-utilization due
to the lack of spectrum reusability consideration.

In recent years, quite a few research efforts have been made for the studies on privacy-preserving spectrum auctions \cite{14chenz}\cite{chen2017secure}\cite{16huangq}\cite{16suny}\cite{13huangq}\cite{11panm}. Most, if not all, of them have focused on privacy preservation for single-sided spectrum auctions \cite{16huangq}\cite{16suny}\cite{13huangq}\cite{11panm}. Different from these works, our work addresses the generic case of double spectrum auctions. There have been a few schemes for privacy-preserving double spectrum auctions \cite{14chenz}\cite{chen2017secure}. But these schemes only addressed privacy issues for one-channel double spectrum auctions. As far as we know, we are the first to consider privacy preservation for multi-channel double spectrum auctions.

\section{Underlying Mechanism and Privacy Goal}\label{sec:Problem Statement}

In this section, we introduce the underlying mechanism of the double multi-channel spectrum auction, and define the cryptographical protocol privacy.

\subsection{TRUE-MCSA Auction Mechanism}\label{sec:true-mcsa}
Consider a single-round double multi-channel spectrum auction where there is a coordinator as the auctioneer, $M$ primary spectrum users as the sellers, and $N$ secondary spectrum users as the buyers. Consider the general case where each seller sells multiple channels, and each buyer requests multiple channels. The auction is sealed-bid and private, and each bidder (seller or buyer) submits its request or bid to the auctioneer by itself, without knowing any information about other bidders' submissions.

More specifically, in the spectrum auction, a seller $m$'s request is denoted by $(s_m,c_m)$ $(s_m > 0 , c_m \geqslant 1)$, meaning that the seller $m$ requires the minimum per-channel payment $s_m$ to sell $c_m$ channels; a buyer $n$'s bid is denoted by $(b_n,d_n) $ $(b_n>0, d_n \geqslant 1)$, representing that the buyer $n$ is willing to pay the maximum price $b_n$ for each channel, and wants to buy at most $d_n$ channels. We call $s_m$ and $c_m$ the seller $m$'s request value and request number; and call $b_n$ and $d_n$ the buyer $n$'s bid value and bid number.

An existing solution to the above-mentioned double multi-channel spectrum auction problem is True-MCSA auction mechanism \cite{13chenz}. We will use True-MCSA as our underlying double multi-channel spectrum auction mechanism. A brief review of True-MCSA auction can be found in appendix A.


\subsection{Cryptographical Protocol Privacy}\label{sec:privacy}

Implicitly, True-MCSA assumes that the auctioneer is trusted. However, if this is not the case, True-MCSA simply leaks all requests and bids to the untrusted auctioneer, and thus no privacy is guaranteed.

To protect the privacy of bidders in the case of an untrusted auctioneer, we introduce an agent to cooperatively perform the auction with the auctioneer. Intuitively, our privacy goal is that as long as the auctioneer and the agent do not collude with each other (one of them may be semi-honest), nothing about the sensitive inputs (i.e., bid values, bid numbers, and request values) of bidders is leaked to them through the auction, except what is revealed from the auction outcome. We formally present this privacy definition as follows.

\begin{mydef}[\textbf{Privacy against semi-honest adversaries}]
Let $f(x,y)$ be a two-party deterministic auction functionality with inputs $x$ and $y$ from the auctioneer and the agent, respectively, and a common auction
outcome $f(x,y)$ for both parties. Suppose that protocol $\Pi$
computes functionality $f(x,y)$ between the auctioneer and the agent.
Let $V^{\Pi}_A(x,y)$ (resp. $V^{\Pi}_B(x,y)$) represent the auctioneer's
(resp. the agent's) view during an execution of $\Pi$ on $(x,y)$. In other
words, if $(x, \textbf{r}^{\Pi}_A)$ (resp. $(y,\textbf{r}^{\Pi}_B)$)
denotes the auctioneer's (resp. the agent's) input and randomness, then
\[
\begin{array}{l}
V^{\Pi}_A(x,y) = (x, \textbf{r}^{\Pi}_A, m_1, m_2,...,m_t),\text{ and} \\
V^{\Pi}_B(x,y) = (y, \textbf{r}^{\Pi}_B, m_1, m_2,...,m_t)
\end{array}
\]
where $\{m_i\}_{i=1}^t$ denote the messages passed between the two parties. Let
$O^{\Pi}(x,y)$ denote the auction outcome after an execution of $\Pi$ on $(x,y)$. Then we have $O^{\Pi}(x,y) = f(x,y)$ for \textbf{correctness}, and say that protocol $\Pi$
\textbf{protects privacy} against semi-honest adversaries if
there exist probabilistic polynomial time (PPT) simulators $S_1$ and
$S_2$ such that
\begin{equation}\label{equ:semisecure1}\small
S_1(x, f(x,y)) \overset{c}{\equiv}
V^{\Pi}_A(x,y)
\end{equation}
\begin{equation}\label{equ:semisecure2}\small
S_2(y, f(x,y)) \overset{c}{\equiv}
V^{\Pi}_B(x,y)
\end{equation}
where $\overset{c}{\equiv}$ denotes computational
indistinguishability.
\end{mydef}

\section{PP-MCSA:Design Challenges and Rationale}\label{sec:challenges}
In this section, we summarize the main challenges in our design, followed by our design rationale to tackle them.
\subsection{Design Challenges}
Recently, some secure mechanisms for double spectrum auctions, such as PS-TRUST or SDSA {\cite{14chenz}\cite{chen2017secure}}, have been proposed. However, they all assumed that in the spectrum auction a seller sells one channel, and a buyer buys one channel, and none of them addressed the privacy preservation issue in double multi-channel spectrum auctions. To protect privacy in double multi-channel spectrum auctions like TRUE-MCSA, we face two challenges indicated as follows.

The first one is the \emph{privacy challenge}. As described in appendix A, TRUE-MCSA involves complex operations in both ``VBG splitting and bidding'' and ``winner determination'' steps. How to perform such operations securely by protecting the sensitive inputs is our first challenge.

The second one is the \emph{efficiency challenge.} Straightforwardly securing the auction in our context may result in heavy overhead and thus may degrade the overall performance. Thus, how to achieve practical efficiency in terms of performance with privacy guarantee consists of our second challenge.



\subsection{Design Rationale}
In order to tackle these two challenges above, we leverage garbled circuits \cite{Yao1982ProtocolsFS}\cite{Lindell2009APO} to carefully design the boolean circuits corresponding to the auction mechanism. Specifically, to achieve privacy, we designate binary flags to indicate various conditions, and implement the auction functionality based on these flags in a data-oblivious way; to achieve efficiency, we carefully cache some intermediate values, so that unnecessary repeated circuits are avoided.

\section{PP-MCSA:Design Details And Proofs}\label{sec:DesignDetail}

In this section, we elaborate our privacy preserving spectrum auction protocol, namely PP-MCSA, and prove that it is secure against semi-honest adversaries.

\subsection{Protocol Framework}

\begin{figure}
  \centering
  \includegraphics[width=0.8\linewidth]{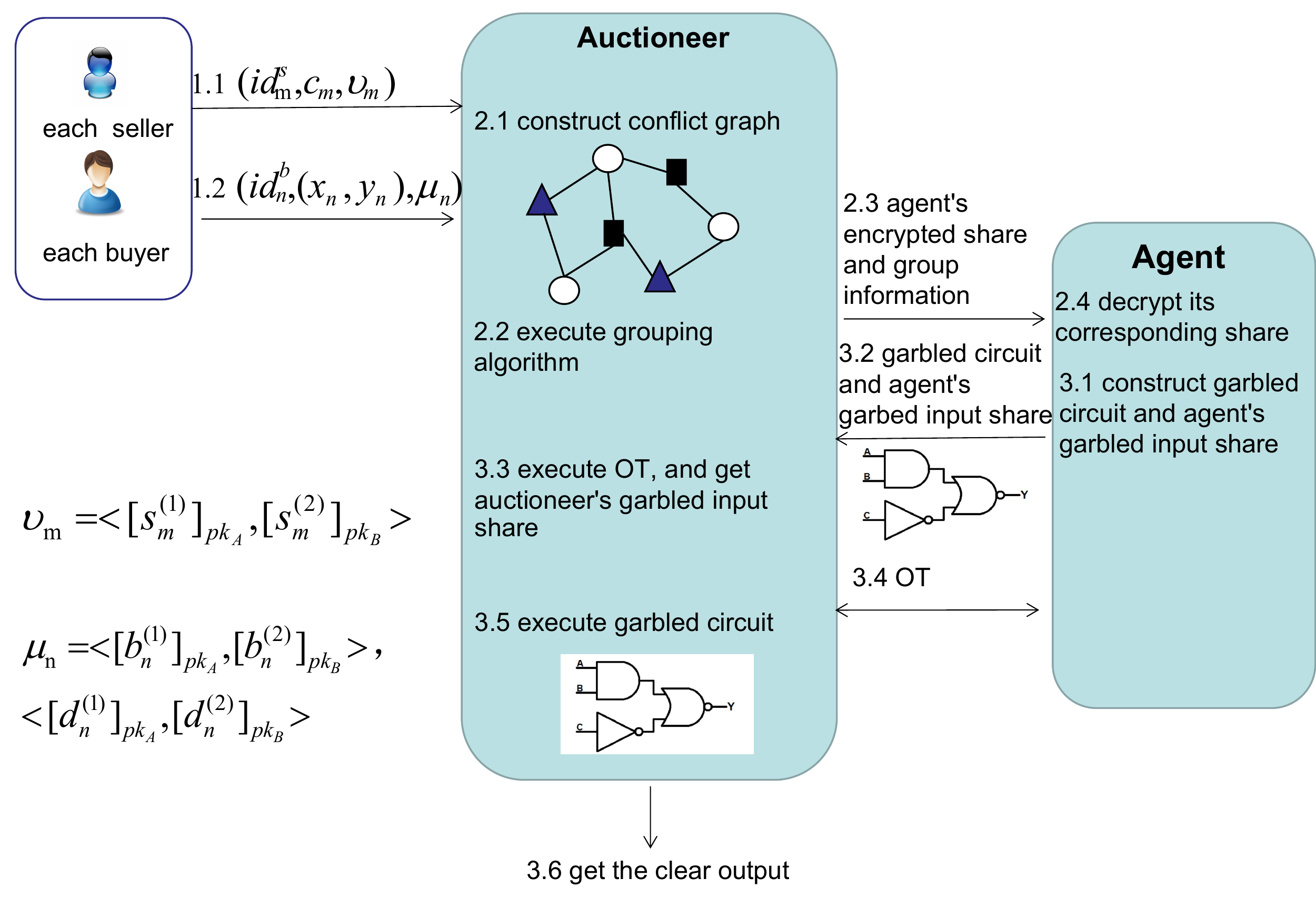}
  \caption{Protocol framework: First, each buyer or seller submits its input with sensitive parts properly split and encrypted; Next, the auctioneer constructs a conflict graph of buyers, executes buyer grouping algorithm and forwards encrypted input shares to the agent; Then, the agent obtains its corresponding input shares by decrypting the encrypted ones, constructs a garbled circuit based on the auction circuit, garbles its input shares, and sends the garbled circuit and garbled input shares to the auctioneer; Finally, the auctioneer obtains its garbled input shares through running an oblivious transfer with the agent, and executes the garbled circuit and outputs the clear result.}\label{fig:AuctionModel}
\end{figure}

In this subsection, we present the protocol framework of PP-MCSA. Generally speaking, PP-MCSA is a secure protocol for double multi-channel spectrum auctions executed between the auctioneer and the agent. We distinguish two types of inputs, \emph{insensitive} and \emph{sensitive} ones, among which the sensitive input needs to be protected in the spectrum auction. We combine public-key encryption with garbled circuits to protect the sensitive input throughout the auction. As shown in Fig.~\ref{fig:AuctionModel}, our protocol consists of three phases, namely, \emph{submission}, \emph{group formation}, and \emph{garbled auction computation}, as specified as follows.


\textbf{Phase \uppercase\expandafter{\romannumeral1}: Submission}

In this phase, sellers and buyers encrypt their respective sensitive inputs, and then send all the necessary inputs to auctioneer. Sensitive inputs include all sellers' request values, all buyers' bid values and bid numbers, while the insensitive inputs include all sellers' IDs and request numbers, and all buyers' IDs and geographic locations. For sensitive inputs, we split all of them into two parts, and then encrypt them respectively with the auctioneer's public key $pk_A$ and the agent's public key $pk_B$. For insensitive inputs, we directly send them to the auctioneer. The tuples that are submitted by sellers and buyers are presented as follows.
\begin{description}
  \item [Seller] $m$$:( id_m^s, \langle [s_{m}^{(1)}]_{pk_{A}},[s_{m}^{(2)}]_{pk_{B}}\rangle,
      c_m)$  for $m=1,2,...,M$
  \item [Buyer] $n$$:(id_n^b, (x_{n}, y_{n}),
      \langle [b_{n}^{(1)}]_{pk_{A}},[b_{n}^{(2)}]_{pk_{B}}\rangle, \langle [d_{n}^{(1)}]_{pk_A}, [d_{n}^{(2)}]_{pk_B}\rangle)$ for $ n=1,2,...,N$
\end{description}
where $[\cdot]_{pk_A}$ and $[\cdot]_{pk_B}$ denote encryptions with $pk_A$ and $pk_B$, respectively, and $x^{(1)} + x^{(2)} = x \pmod{2^{B}}$ for any value $x$, where $B$ is the bit length used.

Additionally, we assume that all communication channels are authenticated and secure, and no one can eavesdrop the data transmitted on the channels.

\textbf{Phase \uppercase\expandafter{\romannumeral2}: Group Formation}

Upon receiving the inputs from sellers and buyers, the auctioneer firstly constructs a conflict graph using all buyers' geographic locations. Then, according to the conflict graph, the auctioneer executes a bid-independent grouping algorithm to divide buyers into different groups, such that any two members of the same group do not conflict with each other.  After group formation, the auctioneer gets  group set $G=\{{G}_{1},{G}_{2},\ldots,{G}_{T}\}$, where the size of group $G_t$ is denoted by $N_t$. An example of group formation is illustrated in step 2.1 in Fig.\ref{fig:AuctionModel}, where nodes represent buyers, edges represent conflict relations between buyers, nodes with the same shape represent members in the same group, and thus three groups are formed. At the end of this phase, the auctioneer sends the agent's encrypted shares of sensitive inputs, and the grouping information to the agent. Then, both the auctioneer and the agent can obtain their respective shares of sensitive inputs by decrypting the corresponding encrypted shares with their public keys.


\textbf{Phase \uppercase\expandafter{\romannumeral3}: Garbled Auction Computation}


In this phase, the agent constructs a garbled circuit based on the auction circuit which we will design in the next subsection, garbles its shares of sensitive inputs, and generates the output decoder which can decode the garbled output. The garbled circuit, the agent's garbled input shares, and the output decoder are then sent to the auctioneer. Upon receiving these data, the auctioneer executes oblivious transfers (OTs) with the agent to get its garbled shares of sensitive inputs. Finally, with both garbled shares of sensitive inputs and the insensitive inputs in hand, the auctioneer computes the garbled circuit to get a garbled auction result, and obtains the clear auction result by decoding the garbled one with the output decoder.

The crux of this phase is to design a boolean circuit for our underlying spectrum auction, True-MCSA. A boolean circuit is in essence the binary representation of a data-oblivious algorithm, whose execution path does not depend on its input. In our case, we only need to design auction algorithms which are data-oblivious for sensitive inputs. In the next subsection, we detail our design of such data-oblivious algorithms.

\subsection{Data-oblivious Auction Algorithms}

In our context, we only need to protect the sensitive inputs from both sellers and buyers. Thus, we only need to perform sensitive input related operations in the garbled circuits. From here on, we represent a garbled $x$ by $[\![x]\!]$, meaning that $x$ needs to be protected and should remain in the garbled form throughout the computations. Our data-oblivious spectrum auction is further composed of four steps as follows.

\textbf{\emph{1) Initialization.}} In our algorithms, we use arrays of tuples to represent both sellers' and buyers' information. Specifically, we use an array of seller tuples $\mathbb{S}$ to represent all sellers, an array of buyer group tuples $\mathbb{G}$ to represent all buyer groups, an array of buyer tuples $\mathbb{G}_t$ to represent all buyers in the group $t$ $(t = 1, \cdots, T)$, and an array of virtual buyer group (VBG) tuples $\mathbb{G}^v_t$ to represent all VBGs derived from the group $t$. The four types of tuples are designed as follows.
\begin{description}{\desclabelstyle{\pushlabel}\desclabelwidth{6em}}
  \item [Seller]tuple: $(id_j^s, s_{j},c_j, w^{s}_{j})$, $j \in [1..M]$
  \item [Group]tuple: $(id_t^g, b^g_t, N_t)$, $t\in [1..T]$
  \item [Buyer]tuple: $(id^b_{t,q}, b_{t,q}, d_{t,q},w^b_{t,q})$, $q \in [1..N_t]$, $t\in [1..T]$
  \item [VBG]tuple: $(id^g_{t},\pi_{t,k}, n_{t,k}, w^v_{t,k})$, $k\in [1..D]$, ${t}\in [1..T]$

\end{description}

In a seller tuple, $id_j^s$, $s_j$ and $c_j$ are the ID, the request value, and the request number of seller $j$, respectively, while $w_j^s$ is a binary flag indicating whether the seller is a winner (1) or not (0). In a group tuple, $id^g_t$, $b^g_t$ and $N_t$ are the ID, the minimum buyer bid, and the size of group $t$. In a buyer tuple, $id^b_{t,q}$, $b_{t,q}$, $d_{t,q}$ are the ID, the bid value, and the bid number of buyer $q$ in the group $t$; $w^b_{t,q}$ describes whether a buyer is  a winner. In a VBG tuple,  $\pi_{t,k}$ and $n_{t,k}$ are the bid, and the size of VBG $k$ derived from group $t$. $w^v_{t,k}$ is a binary flag indicating whether the VBG $k$ is a winning VBG (1) or not (0). Additionally,  $D$ is the maximum bid number of all buyers, which is set as a parameter at the beginning of the auction.

We initialize the arrays $\mathbb{S}$, $\mathbb{G}$, $\mathbb{G}_t$ and $\mathbb{G}^v_{t}$ as follows, where the ``null'' symbol $\bot$ is a placeholder.
\vspace*{-0.5\baselineskip}
\begin{displaymath}
\mathbb{S}=
\left( \begin{array}{rccc}
j: & 1 & \cdots & M \\
id_j^s:& id_1^s & \cdots & id_M^s \\
s_j:& [\![s_1]\!] & \cdots & [\![s_M]\!] \\
c_j:& c_1 & \cdots & c_M \\
w_{j}^s:& 0 & \cdots & 0\\
\end{array} \right),~
\mathbb{G}=
\left( \begin{array}{rccc}
t: & 1 & \cdots & T \\
id_t^g:& id_1^g & \cdots & id_T^g \\
b^g_t:& \bot & \cdots & \bot \\
N_t: &    N_1 & \cdots & N_T \\
\end{array} \right)
\end{displaymath}
\begin{displaymath}
\mathbb{G}_t=
\left( \begin{array}{rccc}
q: & 1 & \cdots & N_t \\
id^b_{t,q}:& id^b_{t,1} & \cdots & id^b_{t,N_t} \\
b_{t,q}:& [\![b_{t,1}]\!] & \cdots & [\![b_{t,N_t}]\!] \\
d_{t,q}:& [\![d_{t,1}]\!] & \cdots & [\![d_{t,N_t}]\!] \\
w^b_{t,q}:&0 & \cdots & 0 \\
\end{array} \right),~
\mathbb{G}^v_t=
\left( \begin{array}{rccc}
k: & 1 & \cdots & D \\
id^g_{t}:& id^g_{t} & \cdots & id^g_{t} \\
\pi_{t,k}:& \bot & \cdots & \bot \\
n_{t,k}:& \bot & \cdots & \bot \\
w^v_{t,k}:& 0& \cdots & 0  \\
\end{array} \right)
\end{displaymath}

\textbf{\emph{2) VBG splitting and bidding.}} In this step, a data-oblivious algorithm should be designed for VBG splitting and bidding. The challenge is that this process depends on both buyers' bid values and their bid numbers, which are sensitive inputs and should be protected.

To design the data-oblivious algorithm, one difficulty is that we do not know the buyers' bid numbers since they are protected in garbled form, and thus we do not know how many VBGs should be derived from each buyer group. To overcome this difficulty, we assume that the maximum bid number $D = \max_{t} D_t$ is known, and hence derive exactly $D$ VBGs from each buyer group. To protect both bid values and bid numbers, we keep them and their related computation results in garbled form, while use appropriate logic circuit to implement all required operations. The resulted algorithm is shown in Algorithm~\ref{alg:VBGSplit}. Note that we only implement MMIN as the VBG bidding method, while GMAX can be similarly implemented.

\floatname{algorithm}{\small Algorithm}
\begin{algorithm}[!htb]
\caption{\small Data-oblivious VBG spplitting and bidding}
\label{alg:VBGSplit}
\begin{algorithmic}[1]\small 
\Require Tuple arrays $\mathbb{G}$ and $\{\mathbb{G}_t\}_{t=1}^T$

\Ensure The tuple array $\mathbb{G}^v_{t}$

\For{$t=1 \rightarrow T$}
\For{$j=1 \rightarrow N_t-1$} \label{line:for1}
        \State $[\![\lambda]\!] \leftarrow ([\![b_{t,j}]\!] < [\![b_{t,j+1}]\!])$;
        \State $swap(\mathbb{G}_t, [\![\lambda]\!], j, j + 1)$; \label{line:swap}
\EndFor \label{line:endfor1}
\For{$k=1\rightarrow D$} \label{line:for2}
\State $[\![n_{t,k}]\!] \leftarrow 0$;\label{line:ntk0}
\For{$j= 1 \rightarrow N_t - 1$}
\State $[\![\gamma]\!] \leftarrow ([\![d_{t,j}]\!] \ge k)$; \label{line:cmp}
\State $[\![n_{t,k}]\!] \leftarrow [\![n_{t,k}]\!] +[\![\gamma]\!]$;\label{line:add}
\EndFor
\State $[\![\pi_{t,k}]\!] \leftarrow [\![b_{t,N_t}]\!]\cdot [\![n_{t,k}]\!]$;\label{line:pi}

\EndFor\label{line:endfor2}
\EndFor

\Return  $\mathbb{G}^v_{t}$
\end{algorithmic}
\end{algorithm}

Some explanations about Algorithm~\ref{alg:VBGSplit} are as follows.

First, for each group $t$, the algorithm compares every pair of neighboring buyer tuples (i.e., tuples $j$ and $j+1$ in $\mathbb{G}_t$ for $j = 1$ to $N_t-1$) in terms of their bid values, and swaps the two tuples if the former is smaller than the later, such that finally the tuple with the minimum bid value is placed at the last position of $\mathbb{G}_t$ (Line~\ref{line:for1} to \ref{line:endfor1}). Note that in Line~\ref{line:swap}, function $swap(\mathbb{G}_t, [\![\lambda]\!], j, j + 1)$ swaps the two tuples $j$ and $j+1$ of $\mathbb{G}_t$ if $\lambda = 1$. For each field $x$ of the tuples, the swapping function can be implemented using the following circuit \cite{13Nikolaenko}:
\begin{align*}
{x'_j} &\leftarrow ((x_j \oplus x_{j+1})\cdot\lambda)\oplus x_j\\
{x'_{j+1}} &\leftarrow {x'_j}\oplus(x_{j}\oplus x_{j+1})
\end{align*}
where $x'_j$ and $x'_{j+1}$ represent the resulted field values. This circuit is very efficient for garbled circuits, since it needs only one non-XOR gate for swapping each pair of bits. Using the free XOR technique, garbled circuits can execute all XOR gates nearly for free, and thus their performances are determined by the number of non-XOR gates executed.

Second, Lines~\ref{line:for2} to \ref{line:endfor2} compute the $D$ VBGs for each group $t$. To compute the $k$th VBG, the bid number of each group member except the last one (who has the minimum bid value) is compared with $k$ (Line~\ref{line:cmp}), and if it is not smaller than $k$, the group member is added to the VBG (Line~\ref{line:add}). Finally, the bid value of the $k$th VBG is computed (Line~\ref{line:pi}).

Note that in the computations, the sensitive inputs, i.e. $b_{t,j}$'s and $d_{t,j}$'s, and their related computation results, i.e., $\lambda$'s, $\gamma$'s, $n_{t,k}$'s and $\pi_{t,k}$'s, are all kept in garbled form, such that the sensitive inputs can be well protected.

\textbf{\emph{3) Winner determination.}} This step applies a variant of McAfee framework to determine winners as shown in Sec.~\ref{sec:Problem Statement}. Since this process contains numerous operations, such as comparisons and selections, depending on requests or bids, designing its data-oblivious version is challenging. In order to address this challenge, our main idea is to introduce some appropriate binary flags to indicate different conditions, and construct suitable circuits based on them to data-obliviously achieve the required functions. We describe the data-oblivious winner determination in Algorithm~\ref{alg:auction1}.

The details of Algorithm~\ref{alg:auction1} are described as follows.

First, both seller tuples and VBG tuples are appropriately sorted as required in McAfee framework (Lines~\ref{line:computeL} to \ref{line:sortVBG}). In Line~\ref{line:computeL}, the total number of selling channels $L$ is computed in the clear, since initially all request numbers $c_j$'s are not protected. In Line~\ref{line:merge}, all VBG tuples from different groups are merged into a uniform VBG tuple array $\mathbb{G}^v = \{id^{v}_k, \pi_k, n_k, w^v_k\}_{k=1}^T$, where $id^{v}_k \in \{id^g_t\}_{t=1}^T$, and then in Line~\ref{line:sortVBG} $\mathbb{G}^v$ is sorted in term of $\pi_k$'s. Note that once sorted (Lines~\ref{line:sortS} \& \ref{line:sortVBG}), all fields of $\mathbb{S}$ and $\mathbb{G}^v$ become garbled, otherwise the ranking information of $s_i$'s and $\pi_k$'s would be leaked.

\floatname{algorithm}{\small Algorithm}
\begin{algorithm}[!htb]
\caption{\small Data-oblivious winner determination}
\label{alg:auction1}
\begin{algorithmic}[1]\small
\Require Tuple arrays $\mathbb{S}$ and $\{\mathbb{G}^v_t\}_{t=1}^T$

\Ensure The winning seller tuple array $\mathbb{W}^s$, the winning VBG tuple array $\mathbb{W}^v$, and the critical request value $\varphi$

\State Compute $L \leftarrow \sum_{i=1}^M c_i$; \label{line:computeL}
\State Sort $\mathbb{S}$ in no-descending order of $s_i$'s, s.t. \label{line:sortS}
\[
[\![s_1]\!]\leq [\![s_2]\!]\leq ...\leq [\![s_M]\!]
\]
\State Merge $\mathbb{G}^v \leftarrow \bigcup_{t=1}^T \mathbb{G}^v_t$;\label{line:merge}
\State Sort $\mathbb{G}^v$ in no-increasing order of $\pi_{k}$'s, s.t. \label{line:sortVBG}
\[
[\![\pi_1]\!]\geq [\![\pi_2]\!]\geq ...\geq[\![\pi_K]\!]
\]
\State $Q \leftarrow \min\{L,K\}$; $[\![\varphi]\!] \leftarrow 0$; $[\![W]\!] \leftarrow 0$;\label{line:computeQ}
\For{$i=1\rightarrow Q$}
\State $[\![\lambda_{M}]\!] \leftarrow 0$; $[\![\delta_{M}]\!] \leftarrow 0$;\label{line:init1}
\State $[\![j_i]\!] \leftarrow 0$; $[\![\varphi_i]\!] \leftarrow  0$; \label{line:init2}
\State $[\![W_i]\!] \leftarrow 0$;\label{line:init3}
\For{$j=M-1\rightarrow 1$}
\State $[\![\lambda_j]\!] \leftarrow [\sum_{l=1}^{j}[\![{c_l}]\!]< i]$; \label{line:lamda}
\State $[\![\delta_j]\!] \leftarrow [\![\lambda_j]\!]\oplus[\![\lambda_{j+1}]\!]$;\label{line:delta}
\State $[\![j_i]\!] \leftarrow [\![j_i]\!] + [\![\delta_{j}]\!]\cdot j$;\label{line:ji}
\State $[\![\varphi_i]\!] \leftarrow [\![\varphi_i]\!] + [\![\delta_{j}]\!]\cdot [\![s_{j+1}]\!]$;\label{line:varphii}
\State $[\![W_i]\!] \leftarrow [\![W_i]\!] + [\![\delta_{j}]\!]\cdot \sum_{l=1}^{j}[\![{c_l}]\!]$;\label{line:Wi}
\EndFor \label{state:end}
\State $[\![\omega_i]\!] \leftarrow [(\sum_{l=1}^i[\![{\pi_l}]\!]) \ge i \cdot [\![\varphi_i]\!]]$; \label{line:omega}
\If{$i > 1$}
\State $[\![\varphi]\!] \leftarrow [\![\varphi]\!] + [\![\varphi_{i-1}]\!] \cdot ([\![\omega_{i-1}]\!]\oplus[\![\omega_{i}]\!] )$; \label{line:varphi}
\State $[\![W]\!] \leftarrow [\![W]\!] + [\![W_{i-1}]\!] \cdot ([\![\omega_{i-1}]\!]\oplus[\![\omega_{i}]\!])$;\label{line:W}
\EndIf

\EndFor\label{line:endforI}
\State $[\![\varphi]\!] \leftarrow [\![\varphi]\!] + [\![\varphi_{Q}]\!] \cdot [\![\omega_{Q}]\!]$;$[\![W]\!] \leftarrow [\![W]\!] + [\![W_{Q}]\!] \cdot [\![\omega_{Q}]\!]$;\label{line:varphiWQ}

\State Reveal $[\![W]\!]$, and $\mathbb{W}^v \leftarrow$ the first $W$ tuples of $\mathbb{G}^v$;\label{line:revealW}
\State Reveal $[\![j_{W+1}]\!]$, and $\mathbb{W}^s \leftarrow$ the first $j_{W+1}$ tuples of $\mathbb{S}$;\label{line:revealjW1}
\State Reveal $[\![\varphi]\!]$ as the critical request value;\label{line:revealVarphi}
\State Resort $\mathbb{W}^v$ in increasing order of $id^g_t$'s, and then in no-increasing order of $\pi_k$'s;\label{line:resort1}
\State Resort $\mathbb{W}^s$ increasing order of $id^s_j$'s;\label{line:resort2}

\Return  $\mathbb{W}^s, \mathbb{W}^v, \varphi$;
\end{algorithmic}
\end{algorithm}

Second, winners are determined with two nested for loops (Lines~\ref{line:computeQ} to \ref{line:endforI}). Specifically, the outer loop iterates over each possible trade $i$, and computes $[\![\omega_i]\!]$ indicating whether trade $i$ is profitable (Line~\ref{line:omega}), the critical request value $[\![\varphi]\!]$ (Line~\ref{line:varphi}), and the number of winning VBGs $[\![W]\!]$ (Line~\ref{line:W}). While the inner loop computes the index $j_i$ of the last winning seller (Lines~\ref{line:init2} \& \ref{line:ji}), the critical request value $\varphi_i$ (Lines~\ref{line:init2} \& \ref{line:varphii}), the number of winning VBGs $W_i$ (Lines~\ref{line:init3} \& \ref{line:Wi}), given trade $i$ is the last profitable trade. Note all these computations are performed in the garbled form.

More specifically, to find the index $j_i$ of the last profitable seller provided the last profitable trade $i$, we introduce two vectors of flags, i.e., $\lambda_j$'s and $\delta_j$'s, where

$\lambda_j$: indicates whether $j \le j_i$, i.e., $\sum_{l=1}^{j}c_l < i$  ($\lambda_j=1$)  or not ($\lambda_j=0$) (Lines~\ref{line:init1} \& \ref{line:lamda}).

$\delta_j$: indicates whether $j = j_i$ ($\delta_j = 1$) or not ($\delta_j=0$) (Lines~\ref{line:init1} \& \ref{line:delta}).

According to the auction logic, the two flag vectors should take values as the following pattern:
\begin{center}
$\left(
\begin{array}{rccccccc}
   j: & 1 & \cdots & j_i-1 & j_i & j_i+1 & \cdots & M\\
    \lambda_{j}: & 1 & \cdots & 1 & 1 & 0 & \cdots & 0 \\
    \delta_{j}: & 0 & \cdots & 0 & 1 & 0 & \cdots & 0 \\
  \end{array}
\right)$
\end{center}

Thus, $\delta_j = \lambda_j \oplus \lambda_{j+1}$ holds (Line~\ref{line:delta}).

With similar idea, we compute the profitable flags $[\![\omega_i]\!]$'s, and the critical request value $[\![\varphi]\!]$ (which is the request value of the critical seller) and the number of winning VBGs $[\![W]\!]$ (Lines~\ref{line:omega} to \ref{line:W}, and Line~\ref{line:varphiWQ}).

It is worth noting that in Line~\ref{line:varphii}, we use $s_{j+1}$ instead of $s_j$, since the critical seller is next to the last winning seller. Additionally, in Lines~\ref{line:lamda}, \ref{line:Wi} \& \ref{line:omega}, for simplicity, we repeatedly use the sum equations of $c_l$'s or $\pi_l$'s. However, in real implementation, it is not necessary to repeatedly compute the sums. Optimally, we can compute each sum just once, and cache them for later use.

Finally, some garbled results are appropriately revealed. Specifically, the number of winning VBGs $[\![W]\!]$ is revealed, and the first $W$ tuples of $\mathbb{G}^v$ form the winning VBG tuple array $\mathbb{W}^v$ (Line~\ref{line:revealW}). Then, $[\![j_{W+1}]\!]$ is revealed as the number of winning sellers, and the first $j_{W+1}$ seller tuples of $\mathbb{S}$ form the winning seller tuple array $\mathbb{W}^s$ (Line~\ref{line:revealjW1}). Next,  $[\![\varphi]\!]$ is revealed as the critical request value(Line~\ref{line:revealVarphi}). At the same time, $\mathbb{W}^v$ and $\mathbb{W}^s$ are appropriately resorted, such that the bid order of winning VBGs from different groups and the request order of winning sellers will not be revealed when decoded in the later (Lines~\ref{line:resort1} \& \ref{line:resort2}). At last, $\mathbb{W}^v$, $\mathbb{W}^s$, and $\varphi$ are returned.

\textbf{\emph{4)Pricing.}} In this step, we compute the selling prices for winning sellers and the buying prices for winning buyers, as described in Algorithm~\ref{alg:Pricing}. Specifically, each winning seller $m$ sells all its $c_m$ channels, and is paid by its selling price $p^s_m = c_m \cdot \varphi$ (Lines~\ref{line:price-for1} to \ref{line:price-endfor1}). Each winning VBG $k$ is charged by its bid $\pi_k$, which is evenly shared by the winning buyers in the VBG. Thus, each winning buyer $n \in G_t$ is charged by its buying price $p^b_n = \min(d_n, D_t) \cdot b^g_t$, where $D_t = \sum_{V \in \mathbb{W}^v} [V.id^v_k = id^g_t]$ is the total number of winning channels for group $G_t$, and $b^g_t = \pi_{t,k}/(n_{t,k} - 1)$ is the minimum bid value of group $G_t$. Note that Lines~\ref{line:price-for2} to \ref{line:price-endfor2} compute the set of winning buyer groups $G^w$, and Lines~\ref{line:price-for3} to \ref{line:price-endfor3} compute the winning buyers in all winning groups and their prices.

\floatname{algorithm}{\small Algorithm}
\begin{algorithm}[!htb]
\caption{\small Pricing}
\label{alg:Pricing}
\begin{algorithmic}[1]\small 
\Require The winning seller tuple array $\mathbb{W}^s$, the winning VBG tuple array $\mathbb{W}^v$, and the critical request value $\varphi$;

\Ensure Winners and their clearing prices;


\For{ $E \in \mathbb{W}^s$}\label{line:price-for1}
\State Reveal $E.id^s_m$;
\State Seller $m$ sells $c_m$ channels, and is paid with $p^s_m \leftarrow c_m \cdot \varphi$;
\EndFor\label{line:price-endfor1}

\State $G^w \leftarrow \phi$;\label{line:price-for2}
\For{ $V \in \mathbb{W}^v$}
\State Reveal $V.id^v_k$ as $id^g_t$;
\State $G^w \leftarrow G^w \cup \{t\}$;
\EndFor\label{line:price-endfor2}
\For{ $t \in G^w$}\label{line:price-for3}
\State $D_t = \sum_{V \in \mathbb{W}^v} [V.id^v_k = id^g_t]$;
\For{$q = 1 \rightarrow N_t - 1$}
\State Reveal $id^b_{t,q}$ as $id^b_n$;
\State $[\![h_t]\!] \leftarrow \min([\![d_n]\!], D_t)$;
\State Reveal $[\![h_t]\!]$ and $[\![b^g_t]\!]$;
\State Buyer $n$ buys $h_t$ channels, and pays $p^b_n \leftarrow h_t \cdot b^g_t$;
\EndFor
\EndFor\label{line:price-endfor3}

\Return All winners and their prices;
\end{algorithmic}
\end{algorithm}

\subsection{Security Analysis}
In this section, we prove that our protocol preserves privacy in the sense of cryptography.

\emph{\textbf{Theorem 1.} As long as the auctioneer and the agent do not collude with each other, PP-MCSA preserves privacy against semi-honest adversaries.}

\textbf{Proof}: The proof of privacy for both Phases I and II is trivial. The reasons are as follows. In Phase I no secure computations are involved, sensitive inputs are secretly shared between the auctioneer and the agent, and hence the view of adversary can be easily simulated.
While in Phase II, group formation is completely dependent on sensitive inputs, and no privacy issues need to be considered. Therefore, we mainly prove the privacy of Phase III.

To prove the privacy of garbled auction computation phase, we actually need to prove the privacy of Algorithms \ref{alg:VBGSplit}, \ref{alg:auction1} and \ref{alg:Pricing} separately, and then by applying the sequential composition theory \cite{04goldreicho} we can conclude that the phase III preserve privacy, and thus the whole protocol also preserve privacy.

We now examine the design of Algorithms \ref{alg:VBGSplit}, \ref{alg:auction1} and \ref{alg:Pricing}. We can see that for every sensitive input related operation, the algorithms compute it in a garbled circuit, and they also store every sensitive input related value by garbled values. At the same time, all garbled values that are revealed in the algorithms carry no more information than what can be inferred from the auction outcome. That is, these algorithms do not revealed any information about the sensitive information except what can be revealed from the auction outcome. By the privacy definition in Section \ref{sec:privacy}, when one party (the auctioneer or the agent) is corrupted, the view of the adversary can be easily simulated (e.g., an encrypted or garbled value can be simulated by a random number of the same bit length). As a result, we can conclude that our algorithms achieve the privacy of garbled circuits, whose privacy is formally proved in \cite{Yao1982ProtocolsFS}.

Therefore, as long as the auctioneer and the agent do not collude with each other, PP-MCSA preserves privacy.
$\boxempty$

\section{Performance Evaluation}\label{sec:experiment}
\subsection{Experimental Setting}

We implement our protocol in two cases: \emph{original} implementation and \emph{improved} implementation. In the original implementation, we implement our algorithms literally, while in the improved implementation, we implement them with cache optimization, where we compute the repeatedly used sums (i.e., Lines~\ref{line:lamda}, \ref{line:Wi} \& \ref{line:omega} in Alg.~\ref{alg:auction1}) just once and cache them for later use. Our experiments are carried out on top of FastGC {\cite{huang2011faster}}, a java-based framework for the garbled circuit computations. We simulate the auctioneer and the agent with two processes on the same computer.
Experimental settings are as follows: buyers are randomly distributed in a $2000\text{m} \times 2000\text{m}$ area, and the interference radius is $400\text{m}$. The request values of sellers and the bid values of buyers are generated randomly in the intervals [1..150] and [1..50], respectively. The both request numbers and bid numbers are generated randomly in the interval [1..10], and thus $D$ is set to 10 which is the maximum bid number. Throughout our experiments, we use bit length $16$, unless otherwise stated, and each point represents the average of 10 times simulation runs.

In the simulation, we run our protocol on a 64-bit Windows 7 Desktop with Intel(R) Core(TM) i5 CPU @3.3GHz and 8GB of memory. We focus on the following two performance metrics:
\begin{compactitem}
\item \emph{Computation overhead}: total running time for executing our protocol by the auctioneer and the agent.
\item \emph{Communication overhead}: total communication cost (data size of all messages that are sent between the auctioneer and the agent).
\end{compactitem}

\subsection{Result Analysis}
We conduct experiments to compare the performance of the original and improved implementations in two cases: (1) when the number of sellers varies; (2) when the number of buyers varies. We further trace the performance of the improved implementation (3) when the bit length of request values and bid values varies; and (4) when bigger numbers of sellers and buyers vary.

\begin{figure}[h]
  \centering
  \includegraphics[width=0.8\linewidth]{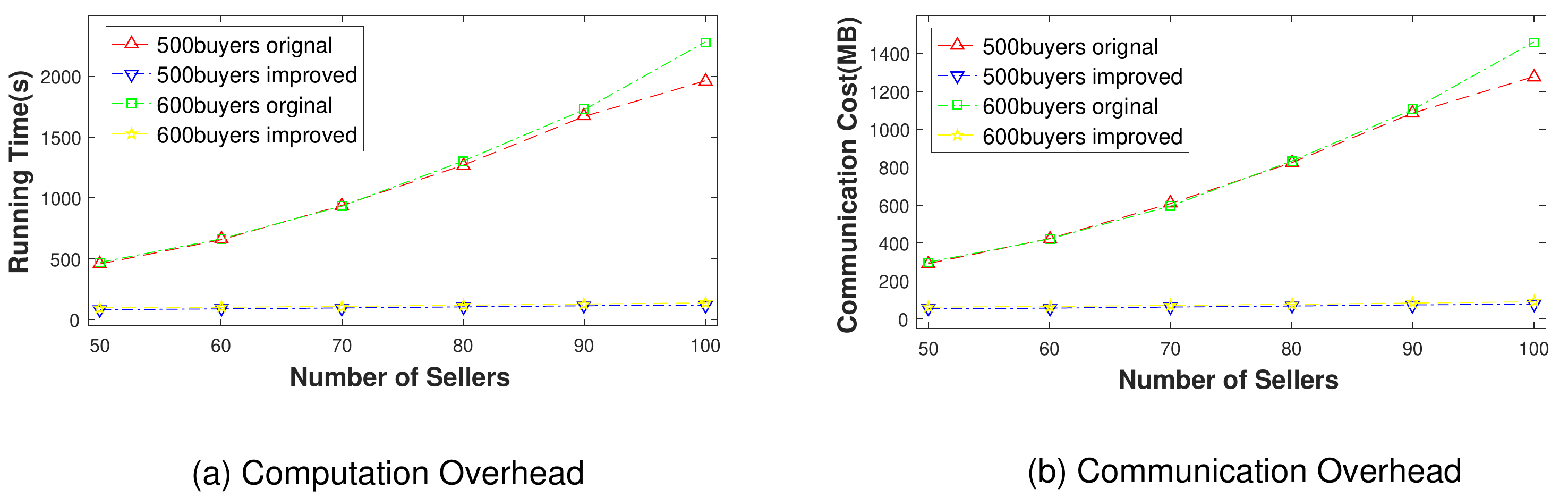}
  \caption{Comparisons of computation and communication overheads between the original and improved implementations as the number of sellers $M$ varies.\protect\\}\label{fig:sellerchange}
\end{figure}

\textbf{(1) Number of sellers varies.} Fig.~\ref{fig:sellerchange} illustrates the comparisons of both computation and communication overheads between the original and improved implementations, when the number of sellers $M$ increases from 50 to 100, and the number of buyers is fixed at $N = 500$, and $N = 600$. We can see that both running time and communication cost of the original implementation increase much faster than those of the improved implementation. The reason is that the cache optimization in the improved implementation (Lines~\ref{line:lamda}, \ref{line:Wi} and \ref{line:omega} in Alg.~\ref{alg:auction1}) avoids repeating the addition computations in the nested loops, and thus greatly reduces the computation and communication overheads.

\begin{figure}[h]
  \centering
  \includegraphics[width=0.8\linewidth]{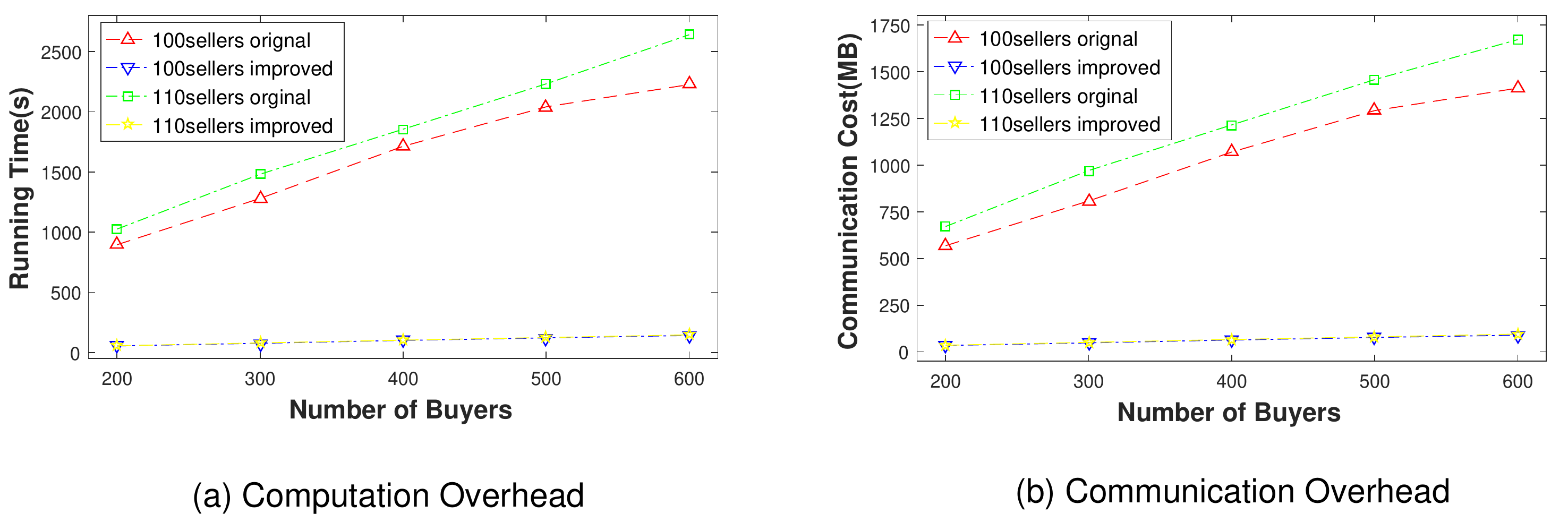}
  \caption{Comparisons of computation and communication overheads between the original and improved implementations as the number of buyers $N$ varies.}\label{fig:buyerchange}
\end{figure}

\textbf{(2) Number of buyers varies.} Fig.~\ref{fig:buyerchange} shows the performance comparisons between the original and improved implementations, when the number of sellers is fixed to $M = 100$ and $M = 110$, and the number of buyers $N$ increases from $200$ to $600$. Similar to Fig.~\ref{fig:sellerchange}, Fig.~\ref{fig:buyerchange} demonstrates that the improved implementation is much more efficient than the original one in term of computation and communication overheads. In the same way, the cache optimization is the source of this performance improvement.

\begin{figure}
  \centering
  \includegraphics[width=0.8\linewidth]{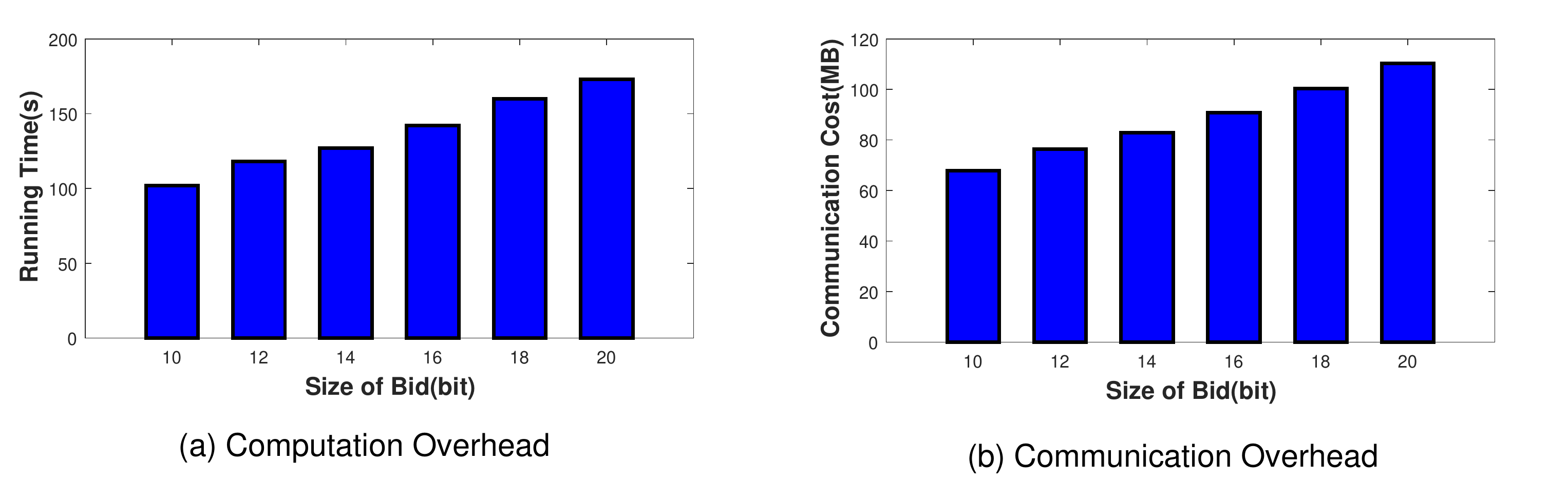}
  \caption{Comparisons of computation and communication overheads between the original and improved implementations as the bit length varies.\protect\\}\label{fig:bitchange}
\end{figure}

\textbf{(3) Bit length varies.} Fig.~\ref{fig:bitchange} traces the impact on performance when changing the bit length of bid values and request values in the improved implementation. We vary the bit length from $10$ to $20$, while fix the number of sellers at $M=80$, and the number of buyers at $N=500$. We can observe that both computation and communication overheads grow linearly as the bit length increases. This is natural, since most of the elemental boolean circuits (e.g., addition, comparison) grow linearly in size when the bit length of its input values increases.

\begin{figure}
  \centering
  \includegraphics[width=0.8\linewidth]{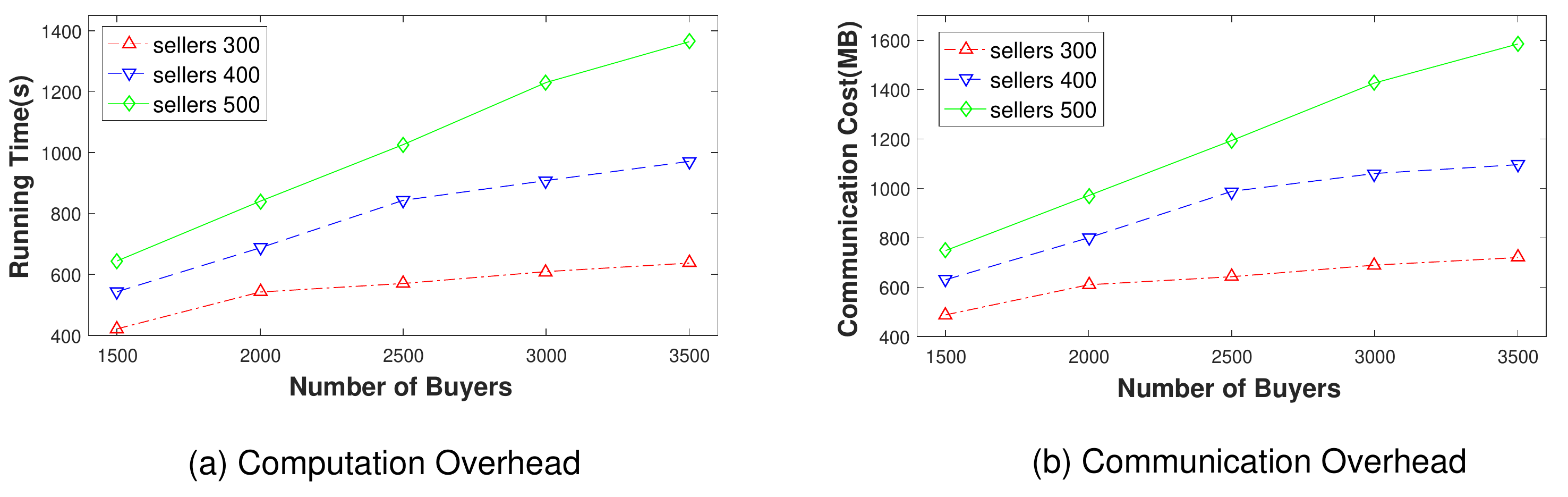}
  \caption{\small Computation and communication overheads of the improved implementation as the big numbers of sellers and buyers vary.}\label{fig:bigchange}
\end{figure}

\textbf{(4) Then bigger numbers of sellers and buyers vary.} Fig.~\ref{fig:bigchange} traces the performance of the improved implementation when the number of buyers varies from $1500$ to $3500$, for the number of sellers $M=300$, $400$ and $500$, respectively. This figure shows that our improved implementation is rather efficient in both computation and communication performance for bigger numbers. For example, all running times are within $23$min, and all communication costs are within $1600$MB. Meanwhile, both computation and communication overheads scale gracefully as the numbers of sellers and buyers increase.

\section{Conclusion}\label{sec:conclusion}

In this paper, we have proposed PP-MCSA, the first privacy-preserving mechanism for multi-channel double spectrum auctions to our knowledge. To address the challenges imposed by the multi-channel double spectrum auction scenario, we have leveraged garbled circuits in our protocol design. Specifically, we design data-oblivious algorithms whose execution path does not depend on their sensitive inputs and then turn these algorithms into garbled circuits to address the privacy challenge. Then, we use cache optimization, which caches some intermediate values to avoid repeated circuits, to improve the garbled circuits and hence address the efficiency challenge. Finally, we have theoretically proved the privacy of PP-MCSA, and experimentally shown that it incurs limited computation and communication overheads.

\section*{Acknowledgment}
The work is supported by the Natural Science Foundation of China under Grant No. 61572031 \& 61502443. We thank the anonymous reviewers for their valuable comments that helped improve the final version of this paper.

\bibliographystyle{IEEEtran}
\bibliography{TAHES}

\appendix
\renewcommand{\appendixname}{Appendix~\Alph{section}}
\section{True-MCSA Auction}
True-MCSA is a truthful double spectrum auction mechanism that allows multi-channel requests from both buyers and sellers, while ensures spectrum reusability. The symbols of the auction can be described in Tab.~\ref{tab:KeySymbolsforTRUE}. Specifically, TRUE-MCSA is composed of the following four steps:

\begin{table}[htbp]\small
  \caption{\small Key Symbols for TRUE-MCSA}
  \vspace*{-0.8\baselineskip}
  \label{tab:KeySymbolsforTRUE}
  \centering
  \begin{tabular}{rl}
    \hline
     $M$, $N$ & numbers of sellers and buyers \\
    $T$ & numbers of buyer groups \\
    $s_{m}$, $c_m$ & seller $m$'s request value and request number \\
    $b_{n}$,$d_n$ & buyer $n$'s bid value and bid number \\
    $(x_{n}, y_{n})$ & location of buyer $n$ \\
    $D_t$ &maximal number of channel of group t\\
    $\pi$ & bid vector of virtual buyer group\\
    $S$ &request vector of sellers\\
    $j(i)$ & the seller in the $i^{th}$ trade\\
    $k_l$ & the last profitable trade\\
    $L$ & sum of sellers channel number\\
     $G$ & $G=\{G_{t}\}_{t=1}^{T}$, global bid set of groups \\
    $G_t$ &the tuple of group t \\
    $G^v_{t}$ &the tuple of virtual buyer group t\\
    \hline
  \end{tabular}
\end{table}

(1) \textbf{Bid-independent Buyer Group Formation}:
In this step, the conflict graph of buyers is constructed in term of their geographic locations, and buyers that do not interfere with each other are grouped into the same group. In this way, buyers in the same group can use the same channels without interference. Note that the group formation algorithm should be bid-independent, otherwise bid manipulation is allowed, and hence the auction becomes untruthful.

(2) \textbf{Virtual Buyer Group (VBG) Splitting and Bidding}:
To address the multi-channel requests from buyers, this step splits a buyer group $G_t$ into $D_t = \max_{i \in G_t}{d_i}$ virtual buyer groups (VBGs), where each VBG only requests for one channel.

After splitting a buyer group into VBGs, we come up with the VBG bidding. Paper \cite{13chenz} proposed two VBG bidding algorithms, member-minimized (MMIN) and group-maximized (GMAX). We only review MMIN as follows. To bid for each VBG derived from a buyer group, the group member with the minimum bid is chosen as the critical buyer, which is removed from all the derived VBGs. Then, each VBG bids with the minimum bid (i.e., the critical buyer's bid) multiplying its size after removing.

(3) \textbf{Winner Determination}: At this point, suppose after VBG splitting and bidding we get totally $K$ VBGs with bid values $\{\pi_k\}_{k=1}^K$. Recall that we have $M$ sellers with request values $\{s_m\}_{m=1}^M$. Then, this step applies McAfee's framework to winner determination as follows.

First of all, the sellers' request values $s_m$'s are sorted in non-decreasing order, and the VBGs' bid values $\pi_k$'s are sorted in non-increasing order:
\begin{center}
$O':s_{1}\leq s_{2} \leq \ldots \leq s_{M}$
\end{center}
\begin{center}
$O'':\pi_{1}\geq \pi_{2} \geq \ldots \geq \pi_{K}$
\end{center}

Then, each seller's request value $s_m$ is rewritten as many times as the number $c_m$ of channels he bid, resulting in the bid mapping between sellers and VBGs as follows:

\begin{displaymath}
\begin{split}
O' :& {\overbrace {s_1\leq {...} \leq s_1}^{c_1}}\leq   {\overbrace {s_2\leq\quad ... \quad \leq s_2}^{c_2}}\, \leq ... \leq {\overbrace {s_M\leq\quad ...\quad \leq s_M}^{c_M}} \\
O'' \!:&{{\pi_{1}\!\geq\! ...\! \geq \!\pi_{c_1}} \!  \geq \! {\pi_{c_1+1}\! \geq \!... \!\geq\! \pi_{c_1+c_2}}\! \geq ... \geq \!{\pi_{\!1+\!{\!\sum^{\!M\!-\!1}_{\!t\!=\!1}c_t}}\!\geq\! ...\! \geq\!  \pi_K \!}}
\end{split}
\end{displaymath}

Finally, find the last profitable trade $k_l$ as:

$k_l = \arg \max_{i\le \min\{L,K\}} \{\sum^i_{t=1}\pi_t \geq i\cdot s_{j(i)}\}$

Here, $L$ represents the total number of channels provided by sellers, namely, $L=\sum_{j=1}^M c_j$, and $j(i)$ computes the seller index $j$ when the trade index is $i$, namely,
\begin{displaymath}
 j(i)=1+\arg \max_{0\le h\le {M-1}}{\{\sum_{t=1}^hc_t<i\}}.
\end{displaymath}

As a result, the last profitable seller is $j(k_l)$. In order to achieve truthfulness, the last profitable seller, as well as all trades involving the seller, should be sacrificed to price the winners. Then the auction winners are the first $j(k_l)-1$ sellers in $O'$ and the first $k=\sum^{j(k_l)-1}_{t=1}{c_t}\le{k_l-1}$ VBGs in $O''$.

(4) \textbf{Pricing}: Each buyer in the same winning VBG pays an even share of the VBG bid, and each winning channel is paid by the price $s_{j(k_l)}$. As a result, each winning buyer pays the sum of what it pays in all the winning VBGs it belongs to, and each winning seller is paid with the product of multiplying its request number and the price $s_{j(k_l)}$.

\end{document}